\documentstyle[12pt,epsf]{article}

\newcommand{\eq}[1]{(\ref{#1})}
\newcommand{\nn}{\nonumber}
\newcommand{\fr}{\frac}

\newcommand{\lsim}{\mbox{ \raisebox{-1.0ex}{$\stackrel{\textstyle <}
{\textstyle \sim}$ }}}

\def\Journal#1#2#3#4{{#1} {\bf #2} (#4) #3}

\def\NPB{{\em Nucl. Phys.} B}
\def\PLB{{\em Phys. Lett.}  B}
\def\PRL{\em Phys. Rev. Lett.}
\def\PRD{{\em Phys. Rev.} D}
\def\ZPC{{\em Z. Phys.} C}
\def\PTP{\em Prog.~Theor.~Phys.}

\begin{document}
\topmargin 0pt
\oddsidemargin 1mm
\begin{titlepage}
\begin{flushright}
 KEK Preprint 98-234\\
 KEK-TH-612\\
 HD-THEP-99-7\\
 October 1999
\end{flushright}

\setcounter{page}{0}
\begin{center}
{\Large \bf  Mass bounds of the lightest CP-even Higgs boson  \\ 
                    in  the two-Higgs-doublet model}
\end{center}
 
\vspace*{6mm}
\begin{center}
{\large \bf  Shinya Kanemura$^\ast$\footnote{
            Address after April 1999 : {\em
            Institut f\"{u}r Theoretische Physik, 
            Universit\"{a}t Karlsruhe, Germany} \\
            \hspace*{0.7cm}(E-mail: kanemu@particle.physik.uni-karlsruhe.de)}, 
             Takashi Kasai$^\dagger$\footnote{
            E-mail: kasai@post.kek.jp}\, 
             and Yasuhiro Okada$^\dagger$\footnote{
            E-mail: yasuhiro.okada@kek.jp} }

\vspace*{6mm}
{\em $^\ast$$^\dagger$ 
       Theory Group, KEK\\
       Tsukuba, Ibaraki 305-0801, Japan}\\
\vspace{2mm}
{\em  $^\ast$ Institut f\"{u}r Theoretische Physik,
              Universit\"{a}t Heidelberg\\  
              Philosophenweg 16, 69120 Heidelberg, Germany}
\end{center}


\begin{abstract}
   The upper and the lower bounds of the lightest CP-even Higgs-boson 
   mass ($m_h$) are discussed in the two-Higgs-doublet model (2HDM)  
   with a softly-broken discrete symmetry.   
   They are obtained as a function of a cut-off scale $\Lambda$ 
   ($\leq 10^{19}$ GeV) 
   by imposing the conditions in which the running coupling constants 
   neither blow up nor fall down below $\Lambda$. 
   In comparison with the standard model (SM),
   although the upper bound does not change very much,    
   the lower bound is considerably reduced.
   In the decoupling regime where only one Higgs boson ($h$) 
   becomes much lighter
   than the others, the lower bound is given, for example, by about 100 GeV   
   for $\Lambda = 10^{19}$ GeV and $m_t = 175$ GeV, 
   which is smaller by about 40 GeV than the corresponding 
   lower bound in the SM.
   In generic cases, 
   $m_h$ is no longer bounded from below by these conditions. 
   If we consider the $b \rightarrow s \gamma$ constraint, 
   small values of $m_h$ are excluded in Model II of the 2HDM.\\ \\
    {\noindent {\it PACS}: 14.80.Cp; 12.60.Fr}
\end{abstract}
\end{titlepage}

   After the discovery of the top quark, the Higgs sector 
   is the last remaining part yet to be confirmed in the  
   Standard Model (SM).  Experimental efforts of 
   {\it Higgs Hunting} are being at a climax 
   in  near future at LEP II, TeV 33 and LHC.  
   Discovery of the Higgs particle is important
   not only in confirming the mechanism of the electroweak 
   gauge-symmetry breaking  
   but also in providing us useful information on physics 
   beyond the SM.  
   Although the mass of the Higgs boson is a free 
   parameter in the minimal SM,
   we can obtain its mass bounds by imposing some 
   theoretical assumptions. 
   If we require the vacuum stability and the validity of perturbation 
   theory below a given cut-off scale $\Lambda$, 
   we can determine the lower and the upper bounds of the Higgs boson mass
   as a function of $\Lambda$, respectively. 
   Allowed region of the Higgs boson 
   and the top quark masses in the SM was examined in ref.~\cite{lindner}.
   For example, for the Plank scale 
   $m_{Pl} \sim 10^{19}$ GeV as $\Lambda$, 
   the lower and the upper bounds become about 145 and 175 GeV at 
   $m_t = 175$ GeV, respectively. 
   This has been reexamined by taking into account the 
   two-loop beta function in ref.~\cite{sm-2loop}. 
   In the minimal supersymmetric standard model (MSSM), on the other hand, 
   the theoretical upper bound on the lightest CP-even Higgs boson mass is
   given by about $120$ GeV for 
   $m_t = 175$ GeV and $m_{\rm stop} = 1$ TeV 
   \cite{oyy,OYYRGE,hhw}.   
   Also, in extended versions of the supersymmetric (SUSY) SM,  
   we can obtain upper bounds, 
   if we demand that all dimensionless coupling constants    
   remain perturbative up to the GUT scale \cite{singlet}. 

   In this letter, we investigate the upper and the lower bounds of 
   the lightest CP-even Higgs-boson mass in the 
   two-Higgs-doublet model (2HDM) with a softly-breaking discrete symmetry
   by requiring the vacuum stability and the validity of perturbation
   theory. By a similar method as used in the SM, we can determine
   these mass bounds as a function of a cut-off scale $\Lambda$.   
   In the 2HDM,  a discrete symmetry is often assumed in order to suppress 
   the flavor changing neutral current (FCNC) 
   in a natural way \cite{gw}. According to the couplings with quarks,
   the 2HDM with such discrete symmetry is classified in two types; namely,   
   one where only one Higgs doublet has Yukawa couplings with the 
   quarks and leptons (Model I), and the other where the one 
   Higgs doublet interacts only with the down-type quarks and 
   leptons and the second 
   one only with up-type quarks (Model II) \cite{hhg}.
   In this letter, we also include soft-breaking terms of the discrete
   symmetry in the Higgs potential.
   Inclusion of these terms does not induce the FCNC problem and may be 
   necessary to avoid the domain wall problem \cite{domain}.
   There have been several works  
   on the Higgs mass bounds in the 2HDM without the soft-breaking term
   \cite{chivkura,kkt,komatsu,sher1,sher2}.
   Our analysis is a 
   generalization of these works to the case with the soft-breaking 
   terms. The results are qualitatively different 
   from the previous works in the region of the large soft-breaking 
   mass, where only one neutral Higgs boson becomes light 
   and the others are much heavier and decouple from the electroweak scale. 
   The lower bound of the lightest Higgs boson mass in this case
   is much reduced in comparison with that in the SM.
   For example for $\Lambda = 10^{19}$ GeV and $m_t = 175$ GeV, 
   while the upper bound is about 175 GeV, which is the almost 
   the same as in the SM,   
   the lower bound is given by 100 GeV. 
   This is considerably smaller than the similar lower bound in the SM 
   which is 145 GeV.
   For the region of the small soft-breaking mass, 
   the lower and upper bounds depend on the soft-breaking mass and 
   there is no longer bounded from below 
   in the case without the soft-breaking mass. In Model II 2HDM
   the constraint from $b \rightarrow s \gamma$ branching ratio excludes 
   the small mass region of the neutral Higgs boson.

The Higgs potential of the 2HDM is given for both Model I and Model II as 
\cite{hhg}  
\begin{eqnarray}
  {V}_{\rm 2HDM}  &=&     m_1^2 \left| \varphi_1 \right|^2 
                          + m_2^2 \left| \varphi_2 \right|^2 - 
                              m_3^2 \left( \varphi_1^{\dagger} \varphi_2 
                                + \varphi_2^{\dagger} \varphi_1 \right) 
                                \nn  
                       + \frac{\lambda_1}{2} 
                               \left| \varphi_1 \right|^4 
                             + \frac{\lambda_2}{2} 
                               \left| \varphi_2 \right|^4 \nn \\
& &                          + \lambda_3 \left| \varphi_1 \right|^2 
                                \left| \varphi_2 \right|^2 
                      + \lambda_4 
                               \left| \varphi_1^{\dagger} \varphi_2 \right|^2
                             + \frac{\lambda_5}{2} 
                             \left\{ 
                               \left( \varphi_1^{\dagger} \varphi_2 \right)^2
                            +  \left( \varphi_2^{\dagger} \varphi_1 \right)^2
                             \right\},  \label{pot}
\end{eqnarray}
where we include the soft-breaking terms for the discrete symmetry.  
For simplicity, we take all the self-coupling constants and the mass 
parameters in \eq{pot} to be real. 
In  Model II $ \varphi_1 $ has couplings
with down-type quarks and leptons and  $ \varphi_2 $ has
couplings with up-type quarks, and only  $ \varphi_2 $
has couplings with fermions in Model I.

 From the above Higgs potential \eq{pot}, it is straightforward to derive
 masses of the Higgs bosons assuming that there is no CP nor charge
 violation at vacuum. Defining the ratio of two vacuum expectation
 values by  
$\tan \beta = \langle \varphi_2 \rangle / \langle \varphi_1 \rangle $,
the masses of the charged Higgs bosons $(\chi^\pm)$ and CP-odd Higgs
boson $(\chi_2)$ are expressed as 
$m_{\chi^\pm}^2 = { M^2} - (\lambda_4 + \lambda_5) v^2 / 2 $ and 
$m_{\chi_2}^2 =  { M^2}  - \lambda_5 v^2$, respectively, 
where $ M = m_3 / \sqrt{\cos \beta \sin \beta}$ and $v \sim 246$ GeV.
The two CP-even Higgs boson masses are obtained by diagonalizing the  
$2 \times 2$ matrix, where each component is given by
$  M_{11}^2 = v^2 \left(\lambda_1 \cos^4 \beta + \lambda_2 \sin^4 \beta 
                    + \frac{\lambda}{2} \sin^2 2 \beta \right)$, $ 
  M_{12}^2 = M_{21}^2 = v^2  \sin 2 \beta 
       \left( - \lambda_1 \cos^2 \beta + \lambda_2 \sin^2 \beta 
                    + \lambda \cos 2 \beta \right)/2$ and  $ 
  M_{22}^2 = v^2 \left(\lambda_1  + \lambda_2  
                    - 2 \lambda \right) \sin^2 \beta \cos^2 \beta 
              + M^2$
where $ \lambda \equiv  \lambda_3 + \lambda_4+ \lambda_5$.
The mass of the lighter (heavier) CP-even Higgs boson $h$ ($H$) 
is then given by 
$m_{h,H}^2 =  \left\{ M_{11}^2 + M_{22}^2 \mp 
\sqrt{(M_{11}^2 - M_{22}^2)^2 + 4 M_{12}^4 }\right\}/2$.
For the case of $v^2 \ll M^2$, they can be expressed by  
\begin{eqnarray}
\!\!\!\!\!\!\!\!  m_h^2 &=& 
v^2 \left(\lambda_1 \cos^4 \beta + \lambda_2 \sin^4 \beta 
                    + \frac{\lambda}{2} \sin^2 2 \beta \right) 
+ {\cal O}(\frac{v^4}{M^2}), \label{h-mass-2hdm} \\
\!\!\!\!\!\!\!\!  m_H^2 &=&  
               { M^2} + v^2 \left(\lambda_1  + \lambda_2  
                    - 2 \lambda \right) \sin^2 \beta \cos^2 \beta
+ {\cal O}(\frac{v^4}{M^2}). 
\end{eqnarray}
Notice that the free parameter $M$ characterizes properties of
the Higgs bosons in this model. In the case of $M^2 \gg \lambda_i v^2$, 
the masses of all the Higgs bosons but $h$ 
become close to $M$.  In this region, these heavy Higgs 
bosons decouple from the low-energy observable 
due to the decoupling theorem \cite{appel} and below the scale $M$  
the model is effectively regarded as the SM with one Higgs 
doublet. On the other hand, if $M^2 \sim  \lambda_i v^2$, the masses are 
controlled by the self-coupling constants, and thus the heavy Higgs 
bosons do not decouple and the lightest CP-even Higgs boson can 
have a different property from the SM Higgs boson \cite{kanemura}.

Let us discuss the conditions for validity of perturbation
theory and the vacuum stability.
For the first condition, we require that 
the running coupling constants of the Higgs self-couplings and the Yukawa 
couplings do not blow up below a certain energy scale $\Lambda$; 
\begin{eqnarray}
  \forall  \, \lambda_i(\mu) < 8 \pi, \; y_t^2(\mu) < 4 \pi \,, \label{blowup}
\end{eqnarray}
for a renormalization scale $\mu$ less than $\Lambda$.  
For the requirement of the vacuum stability,  
we assume that the quartic interaction 
terms in the potential do not give negative contribution 
for all directions of scaler fields at each energy scale 
up to $\Lambda$. This condition leads to 
\begin{eqnarray}
&& \;\;\;\; \lambda_1(\mu) > 0, \;\; \lambda_2(\mu) > 0, \nn \\
&& \sqrt{\lambda_1(\mu) \lambda_2(\mu)} + \lambda_3(\mu) + 
    \rm{min} \left[ 0,\lambda_4(\mu) + \lambda_5(\mu), 
               \lambda_4(\mu) - \lambda_5(\mu) \right]\, >\, 0 ,\label{stable}
\end{eqnarray}
for $\mu < \Lambda$. We also require that the tree-level
Higgs potential at the weak scale does not have any global minimum 
except for the one we consider.
In particular, we assume that there is no CP nor charge breaking at the
global minimum
\footnote{
The vacuum stability condition here is slightly different from 
that in ref.~\cite{sher2}, where   
they have put $\lambda_4(\mu) + \lambda_5(\mu) < 0$ and 
$\lambda_5(\mu) < 0$ below $\mu < \Lambda$ in addition to \eq{stable} 
in the model with $M^2 = 0$.   
In the case of $M^2 \sim 0$, 
our condition is essentially the same as that in ref.~\cite{sher2},
because we then have 
$\lambda_4 + \lambda_5 < 0$ and $\lambda_5 < 0$ at the electroweak scale 
from the positiveness of the squared-masses of $\chi^\pm$ and $\chi_2$  
and we can show that 
these inequalities tend to be preserved at higher energy scale
according to the 2HDM RGE's.
}.
These conditions imposed on the coupling constants 
at a high energy region are transmitted into constraints 
on the coupling constants at the electroweak scale
and then on the masses of Higgs bosons.

In the decoupling case where $M^2 \gg \lambda_i v^2$, 
there can be a sizable correction on the lightest
CP-even Higgs boson mass at a low energy scale.
In order to include this effect, instead of calculating 
the $\lambda_i $'s at the weak 
scale from the RGE and using the tree-level mass formulas,
we adopt the following procedure. 
We determine the $\lambda_i$ at the scale $M$ 
by using the 2HDM RGE in the region between $\Lambda$ and $M$,    
and then calculate the CP-even Higgs boson mass according to
the tree-level formulas. 
Since the effective theory below  $M $ is
just the SM with one Higgs doublet, 
we use the SM RGE from  $M $ to $m_h$
to evaluate the lightest Higgs boson mass.
Although this procedure is not
really justified for $M^2 \sim  \lambda_i v^2$,  
we calculate the mass in this way 
because the correction from 
the SM RGE is numerically very small in such case.

In our analysis, we use the 1-loop RGE's for the SM and the 2HDM 
which are found, for example, in ref.~\cite{komatsu,inoue}. 
We only consider the top-Yukawa coupling contribution as the Yukawa 
interaction.  
The running top mass is defined as  
$\overline{m_t}(\mu)=\frac{1}{\sqrt{2}} y_t(\mu) v \sin{\beta}$
 and it is related to the pole mass $m_t$ by 
 $\overline{m_t}(m_t)=m_t(1-\frac{4}{3\pi}\alpha_s(m_t))$. 
   There are important phenomenological constraints on the 
   2HDM. From the low-energy electroweak precision tests,  
   the $\rho$ parameter should be closed to unity, which means that    
   the custodial $SU(2)_V$ symmetry should not be badly broken 
   in the Higgs sector.
   We evaluate the 2HDM contribution to the $\rho$ parameter according to 
   refs.~\cite{rho}. Taking account of the experimental data up to 95\% CL 
   \cite{hhm}, we here set the condition 
   $\Delta \rho_{\rm 2HDM} = - 0.0020 - 0.00049 \frac{m_t - 175 {\rm GeV}}
        {5 {\rm GeV} } \pm 0.0027$
   for our analysis, where $\Delta \rho_{\rm 2HDM}$ is the 
   extra contribution of the 2HDM to the $\rho$ parameter\footnote{
   We here set the reference value of the SM Higgs mass into 100 GeV.  
   We also include uncertainties from the strong coupling constant
   and the electromagnetic coupling constant at the Z pole  
   for our evaluation of the $\rho$ paremeter.}. 
%
%
   Another experimental constraint is obtained from the 
   $b \rightarrow s\gamma$ measurement \cite{bsg-exp}. 
   It is known that there is very strong constraint on the
   charged-Higgs boson mass from this process in the case of  
   Model II,  while Model I is not strongly constrained.
   We calculate the $b \rightarrow s\gamma$ 
   branching ratio with the next-to-leading order
   QCD correction~\cite{bsg-2hdm} and use its constraint to
   determine the allowed region of the parameter space.

In the actual analysis,  we first fix parameter sets
of $m_h$, $\tan \beta$ and $M$. Since the Higgs potential
contains three masses and five coupling constants,  
 the number of free parameters is 
four  with fixing $v= 246$ GeV for each set of the parameter choice. 
We examine four-dimensional parameter space of $\lambda_1$, 
$\lambda_2$,  $\lambda_4$ and  $\lambda_5$ under the experimental 
constraints above and obtain a maximum scale $\Lambda$ 
where one of the conditions \eq{blowup} and \eq{stable} is broken. 
We also put $m_Z = 91.19$ GeV and $\alpha_S(m_Z) = 0.118$.
The mass of the top quark is fixed as 175 GeV in our main analysis 
and later the dependence on $m_t$ is discussed.  

Let us first consider the case of the decoupling regime ($v^2 \ll M^2$). 
All the Higgs bosons but $h$ are all heavy and 
their masses are almost degenerate around $M$. 
Fig.~1 shows that the contour plot of each $\Lambda$
($ = 10^{19}, 10^{16}, 10^{13}, 10^{10}, 10^{7}, 10^{4}$ GeV) 
for $M = 1000$ GeV on the $m_h$-$\tan \beta$ plane.  
The $\tan \beta$ dependence is not so sensitive except for the 
small $\tan \beta$ region where the top-Yukawa coupling constant 
blows up at a low energy scale. For the smaller values of 
$m_h$, $\lambda_2$ tends to become negative because of 
the negative effect of $y_t^4$-term 
in the RGE for $\lambda_2$.
On the other hand, for a large value of $m_h$, 
$\lambda_2$ blows up at a low energy scale.    
There is no difference between Model I and Model II in the decoupling regime, 
because the constraint from $b \rightarrow s \gamma$ is not important in 
this case.

 The qualitative result may be understood by looking at the RGE's.  
 From eq.~\eq{h-mass-2hdm}, 
$m_h^2$ is approximately given by $\lambda_2 v^2$ for $tan \beta \gg 1$,
and the RGE for $\lambda_2$ is given by
\begin{equation}
16 \pi^2 \mu \frac{d \lambda_2}{d \mu}  =  
 12 \lambda_2^2  
 - 3 \lambda_2 (3 g^2 + g'^2) 
 + \fr{3}{2} g^4 + \fr{3}{4} (g^2 + g'^2)^2    
+ 12 \lambda_2 y_t^2 - 12 y_t^4 + A,   \label{lam2-approx} 
\end{equation}
where $A = 2 \lambda_3^2 + 
 2 (\lambda_3 + \lambda_4)^2 + 2 \lambda_5^2 > 0$.
 When we fix the coupling normalization by 
$m_H^{SM}=\sqrt{\lambda_{SM}} v$, 
the SM RGE for $\lambda_{SM}$ is  obtained by
substituting $\lambda_{SM}$ and  $y_t^{SM}$ to $\lambda_2$ and $y_t$ in 
eq.~\eq{lam2-approx} and neglecting the $A$ term in the RHS. Thus 
the difference is only in the existence of the positive 
term $A$ in eq.~\eq{lam2-approx}. 
This term works to improve the stability of vacuum to some extent, and   
the lower bound is expected to be reduced in the 2HDM.

Next we see the case of the mixing regime ($M = 100$ GeV $\sim m_z$), 
where the heavy Higgs masses are realized only by the 
large $\lambda_i$'s ($i = 1 - 5$) and their mixing. 
In this case, the data from the low energy experiment 
strongly constrain the model. 
The contour plots for each $\Lambda$ on $m_h$-$\tan \beta$ plane 
in Model I and Model II are shown in figs. 2(a) and 2(b), 
respectively. 
We can see in figs.~2(a) and 2(b) that there is an allowed region 
for $\Lambda = 10^{19}$ GeV in Model I, 
while the largest $\Lambda$ is less than $10^{4}$ GeV in Model II 
because the $b \rightarrow s \gamma$ measurement 
gives a strong constraint for Model II 2HDM.
Note that the allowed region in fig.~2(a) 
lies around $m_h \sim m_Z$ ($\sim M$) for large $\tan \beta$. 
This is because that, in the region of $M^2 < \lambda_2 v^2$, 
the mass of the lighter CP-even Higgs boson $h$ comes from $M_{22} \sim M$ 
and the heavier Higgs boson $H$ has the mass of 
$M_{11} \sim \sqrt{\lambda_2} v$.  
On the other hand, in the decoupling regime, the situation is reversed and 
the $h$ boson has the mass of $M_{11} \sim \sqrt{\lambda_2} v$.          

We repeated the above analysis for various values of $M$ 
and obtained the upper and lower bounds of the lightest 
CP-even Higgs boson masses for various cut-off scales,
which are shown in the contour plots 
in the $m_h$-$M$ plane in figs.~3, (a) and (b) for 
Model I and II, respectively.  
In fig~3(a), the qualitative behavior of the allowed region is 
understood from the above argument on the mass matrix.
For the region of $M^2 \ll \lambda_2 v^2$, the allowed region of $m_h$ 
lies around $m_h \sim M$, 
and that becomes along $\sqrt{\lambda_2} v$ and 
no longer depends on $M$ for $M^2 \gg \lambda_2 v^2$.
Though there are the upper bounds of $m_h$ for each $\Lambda$,  
$m_h$ is not bounded from below by our condition.  
Our results at $M = 0$ are consistent to those in \cite{sher2}.  
If we take account of the experimental result of $b \rightarrow s \gamma$,
$m_h$ is bounded from below in the case of Model II as seen in fig.~3(b)  
because the small $M$ region ($M \lsim 350$ GeV) necessarily corresponds to
the light charged Higgs boson mass and is excluded by 
the $b \rightarrow s \gamma$ constraint\footnote{
For the estimation of theoretical uncertainties we added in quadratures 
the errors form the various input parameters.
If we use more conservative way to add theoretical uncertainties for the 
$b \rightarrow s \gamma$ evaluation, the bound on the charged Higgs boson 
or on the $M$ in Model II becomes rather smaller\cite{bsg-2hdm}. 
The lower bound of $m_h$ due to the $b \rightarrow s \gamma$ constraint  
is then reduced by a few GeV according to the change of the allowed 
region of $M$.
}.

Finally, we show the figure in which 
the results in the SM and the 2HDM (Model I and II) 
are combined on $m_h$-$M$ plane (fig.~4). 
For a reference, the upper and lower bounds of the lightest 
CP-even Higgs mass in the MSSM are also given for the case that
the stop mass is 1 TeV. 
These lines are calculated by a similar method described 
in ref.~\cite{OYYRGE}: 
namely we use the SUSY relation for Higgs self-coupling constants
at the 1 TeV scale and use the 2HDM RGE between 1 TeV and
$M$, and the SM RGE between $M$ and $m_h$ scale. In this figure,  
$M$ is the CP-odd Higgs boson mass in the case of the MSSM. 
It is easy to observe from this figure that  
the difference of the bounds among the SM, the 2HDM(I) and the 2HDM(II).  
We here choose, as an example,  
$\Lambda = 10^{19}$ GeV for the results in the SM and the 2HDM 
at $m_t = 175$ GeV.  
While the upper bounds in these models are all around 175 GeV, 
the lower bounds are completely different; 
about 145 GeV in the SM, about 100 GeV in the Model II and no bound in 
Model I. 

In order to see the top quark mass dependence of the above results, 
we have repeated the analysis for $m_t= 170$ GeV and $180$ GeV. 
It turns out that the lower bound has sizable dependence
of the top mass whereas the upper bound does not change very much.
For example, the lower line for $\Lambda = 10^{19}$ GeV
in the 2HDM shown in fig.~4 shifts to lower (upper) by
9 GeV for $m_t= 170$ $(180)$ GeV at $M = 1000$ GeV, 
but the corresponding shift for the upper line is about 3 (4) GeV. 
In Table 1, 
we list the $m_t$ dependence of the lightest CP-even Higgs mass bounds 
for each value of $\Lambda$ in the SM and  
the 2HDM for $M = 1000$ GeV and for 
$M = 200$  GeV (Model I).


   We also comment on a question    
   how much our results are improved if a higher order 
   analysis is made in the effective potential method. 
   In the SM, the next-to-leading order analysis of 
   the effective potential shows that the lower bound reduces 
   by about 10 GeV ($\Lambda = 10^{19}$ GeV) \cite{sm-2loop}.   
   It may be then expected that a similar reduction of the lower bound  
   would occur in the 2HDM by doing such higher order analysis.

   We have analyzed the upper and the lower bounds of the lightest
   CP-even Higgs boson mass in the 2HDM with a softly-broken discrete 
   symmetry 
   by requiring that the running coupling constants neither blow up nor 
   fall down below $\Lambda$. 
   While the upper bound has been found to be almost the same as in SM, 
   the lower bound turns out to be much reduced. In particular 
   in the decoupling regime, both Model I and Model II give the lower bounds
   of about 100 GeV for $\Lambda = 10^{19}$ GeV, 
   which is lower by $40$ GeV than the SM result. 
   In this regime,  the properties of the lightest  
   Higgs boson such as the production cross section and the decay branching
   ratios are almost the same as the SM Higgs boson. 
   In this letter, we have not explicitly considered constraint from the 
   Higgs boson search at LEP II, but  
   if the Higgs boson is discovered with the mass around $100$ GeV 
   at LEP II or Tevatron experiment
   in near future and its property is quite similar to the
   SM Higgs boson, the 2HDM with very high cut-off scale is another
   candidate of models which predict such light Higgs boson 
   along with the MSSM and its extensions.

The authors would like to thank Kenzo Inoue for useful comments. 
The work of S.K. was supported in part by the Japan Society for 
the Promotion of Science (JSPS).
The work of Y.O. was supported 
in part by the Grant-in-Aid of the Ministry of Education, Science, 
Sports and Culture, Government of Japan (No.09640381), Priority area 
``Supersymmetry and Unified Theory of Elementary Particles'' (No. 707), 
and ``Physics of CP Violation''
(No.09246105). 
S.K. also would like to thank Otto Nachtmann for his warm hospitality 
at the Institute for Theoretical Physics in the University of Heidelberg.

\newpage

\section*{Table Caption}

\newcounter{TAB}
\begin{list}{{\bf Table \arabic{TAB}}}{\usecounter{TAB}}
\item
A list of 
the lower and upper bounds of the lightest CP-even Higgs mass in GeV  
for each $m_t$ ($= 170, 175, 180$ GeV) and $\Lambda$ 
($= 10^{19},10^{16},10^{13},10^{10},10^{7},10^{4}$ GeV)
in the SM as well as the 2HDM 
for $M = 1000$ GeV and for $M = 200$ GeV (Model I).     
Model I and II give the same bounds for $M = 1000$ GeV.
\end{list}

\section*{Figure Captions}

\newcounter{FIG}
\begin{list}{{\bf FIG. \arabic{FIG}}}{\usecounter{FIG}}
\item
The allowed region of the lightest CP even Higgs boson mass 
as a function of $\tan {\beta} $ for different values of the 
cut-off scale $(\Lambda)$ for $M = 1000$ GeV in the 2HDM. 
The top mass is taken to be 175 GeV. For each  
$\Lambda$ ($= 10^{19},10^{16},10^{13},10^{10},10^{7},10^{4}$ GeV)
the inside of the contour is allowed. There is no difference 
between  Model I and Model II in this figure.
\label{fig:m1000}
\item
The allowed region of the lightest CP even Higgs boson mass 
as a function of $\tan{\beta}$ for 
different values of $\Lambda$ for $M = 100$ GeV
in the Model I (a) and Model II (b) 2HDM. 
The top mass is taken to be 175 GeV. 
For the Model II lines for $\Lambda = 1000$ and $3000$ GeV
are shown.\label{fig:m100}
\item
The upper and lower bounds of the lightest CP even Higgs 
boson mass as a function of $M$ for different values of $\Lambda$
in the Model I (a) and Model II (b) 
2HDM for $m_t=175$ GeV.
\label{fig:t175}
\item
The upper and the lower bounds of the lightest CP even Higgs 
boson mass in the Model I and II 2HDM and the SM Higgs boson mass
for $\Lambda=10^{19}$ GeV. The upper and lower bounds of the 
lightest CP even Higgs boson mass in the MSSM are also shown for
the case that stop mass is 1 TeV. In this case $M$ corresponds to
the CP-odd Higgs boson mass in the MSSM.\label{fig:combine}\\
\end{list}

\begin{center}
\begin{tabular}{|c|c||c|c|c|}\hline
&  $\Lambda$  (GeV)& 
   $ m_t = 170$ GeV & $m_t = 175$ GeV & $m_t = 180$ GeV \\  \hline\hline 
Standard Model 
&    & 133 - 172  & 143 - 175 & 153 - 179\\   
2HDM ($M=1000$GeV)& $10^{19}$ 
     &  93 - 172 & 102 - 175 & 111 - 179 \\
2HDM I  ($M=200$GeV)&& 79 - 171 & 84 - 175 & 91 - 179   \\ \hline
& & 133 - 180 & 142 - 182 & 152 - 186\\   
& $10^{16}$ 
  &  89 - 180  & 96 - 183 & 104 - 186 \\
& &  73 - 179 & 80 - 182 &  85 - 185\\ \hline
& & 132 - 192 & 141 - 194 & 150 - 197 \\   
& $10^{13}$ 
  &  85 - 193 & 90 - 195 & 97 - 197 \\
& &  68 - 191 & 72 - 193 & 77 - 195 \\ \hline
& & 129 - 215 & 138 - 216 & 147 - 217\\   
& $10^{10}$
  &  85 - 216 & 89 - 216 & 93 - 218\\
& &  64 - 208 & 67 - 208 & 70 - 207\\ \hline
& & 122 - 264 & 130 - 264 & 138 - 264  \\   
& $10^{7}$
  &  84 - 266 & 88 - 266 & 93 - 265 \\
& &  64 - 238 & 67 - 241 & 69 - 241 \\ \hline
& & 101  - 460   & 107 - 458 & 113 - 458\\   
& $10^{4}$
  &  84 - 480 & 88 - 480 & 92 - 478 \\
& &  63 - 343 & 66 - 342  & 68 - 342 \\ \hline
\end{tabular} 
\end{center}

\vspace{6cm}

\begin{center}
{\Large \bf Table 1} 
\end{center}


\pagestyle{empty}

\def\EPSDIR{}
\def\EPSSCALE{0.90}
\def\fnum@figure{FIG.~\thefigure}

\begin{figure}[htbp]
\begin{center}
\makebox[0cm]{
\epsfbox{\EPSDIR 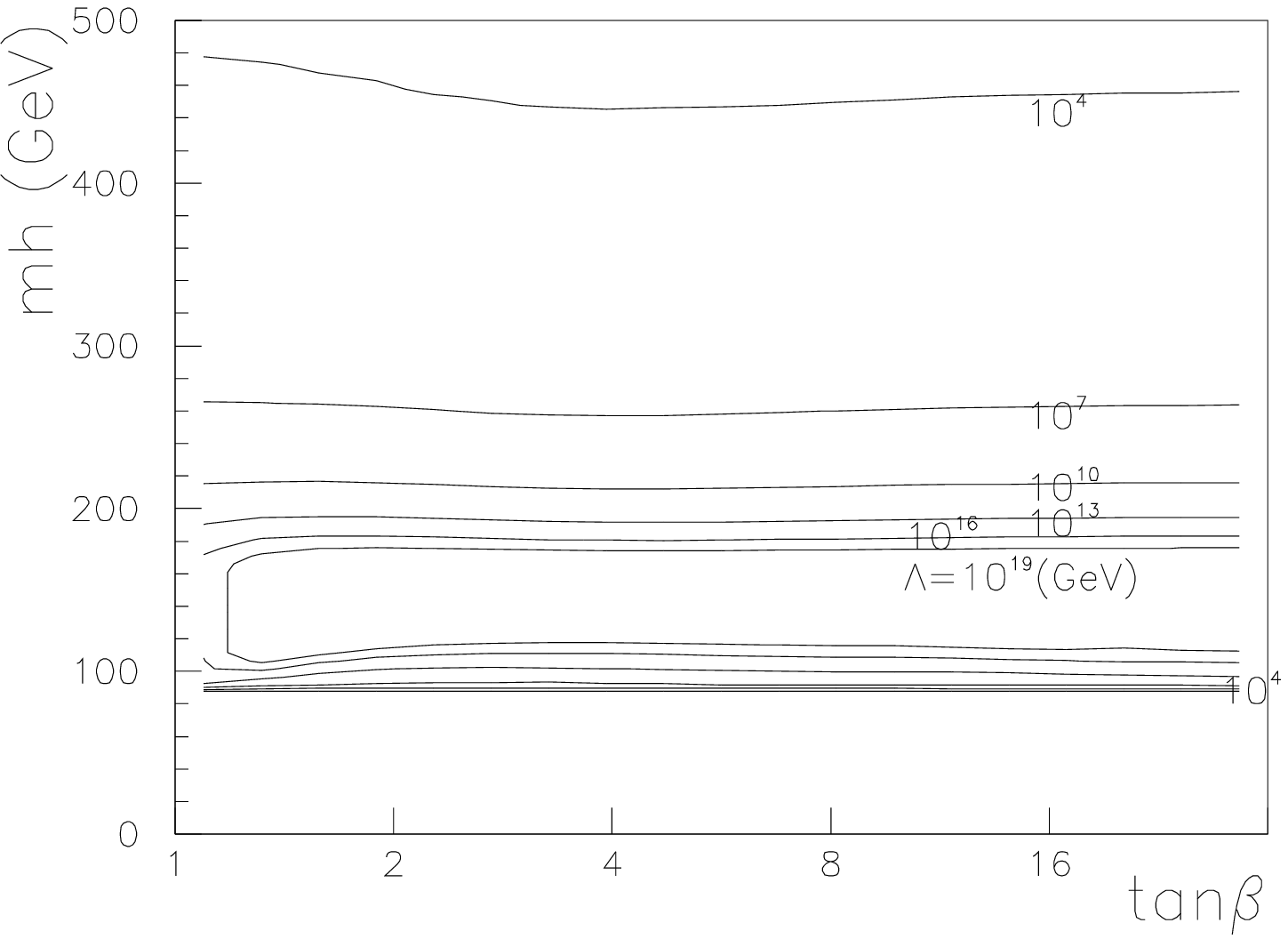 }
}\\ 
{\Large\bf Fig.~\ref{fig:m1000}}
\end{center}
\end{figure}
\clearpage


\begin{center}
\begin{tabular}{cc}
\parbox{15cm}{
\epsfbox{\EPSDIR 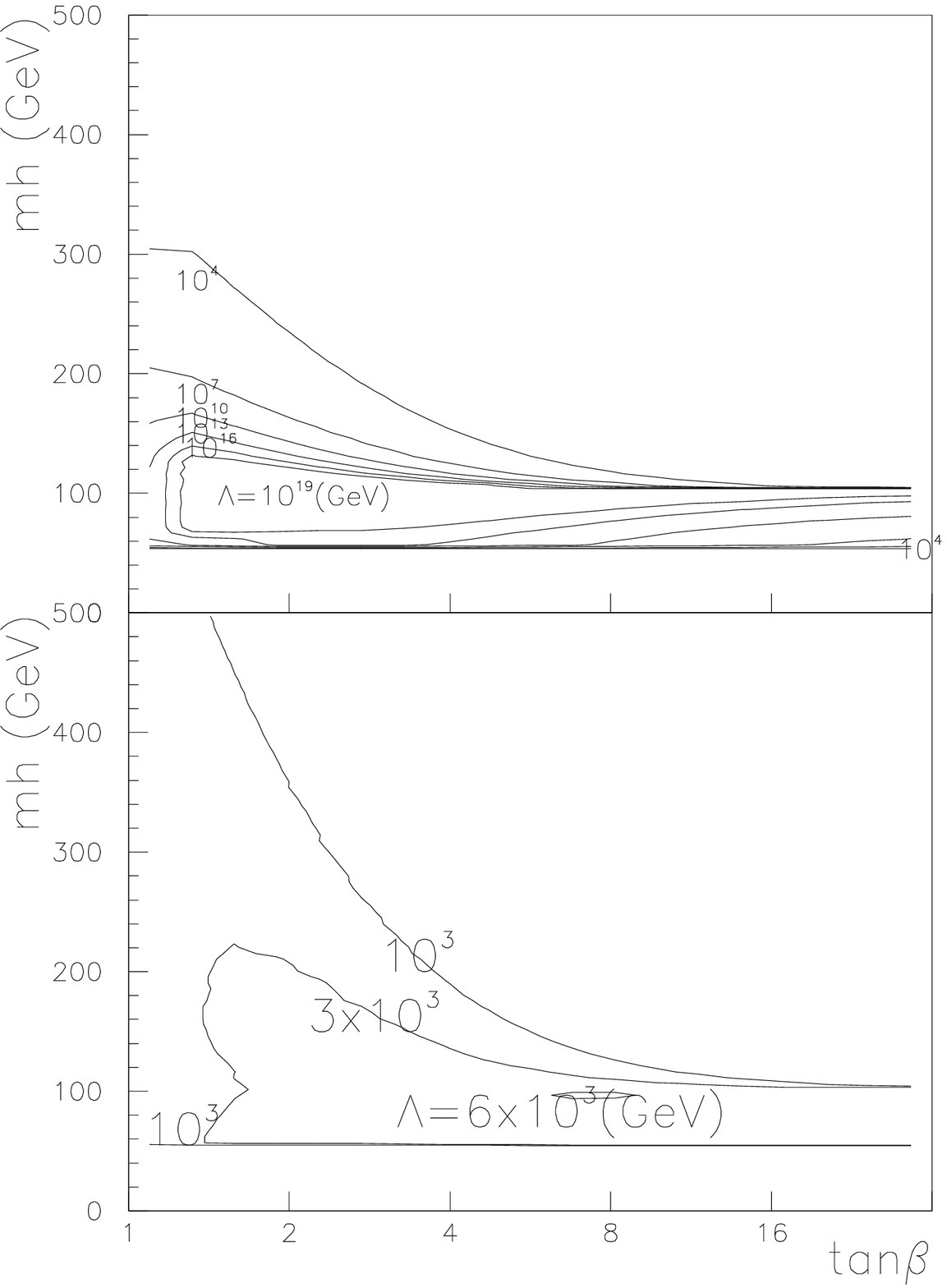}} &
\begin{tabular}{c}
\makebox(5,250){(a)} \\ \makebox(5,250){(b)}  \end{tabular}
\end{tabular}
\vfill
{\Large\bf Fig.~\ref{fig:m100}}
\end{center}
\clearpage

\begin{center}
\makebox[0cm]{
\epsfbox{\EPSDIR 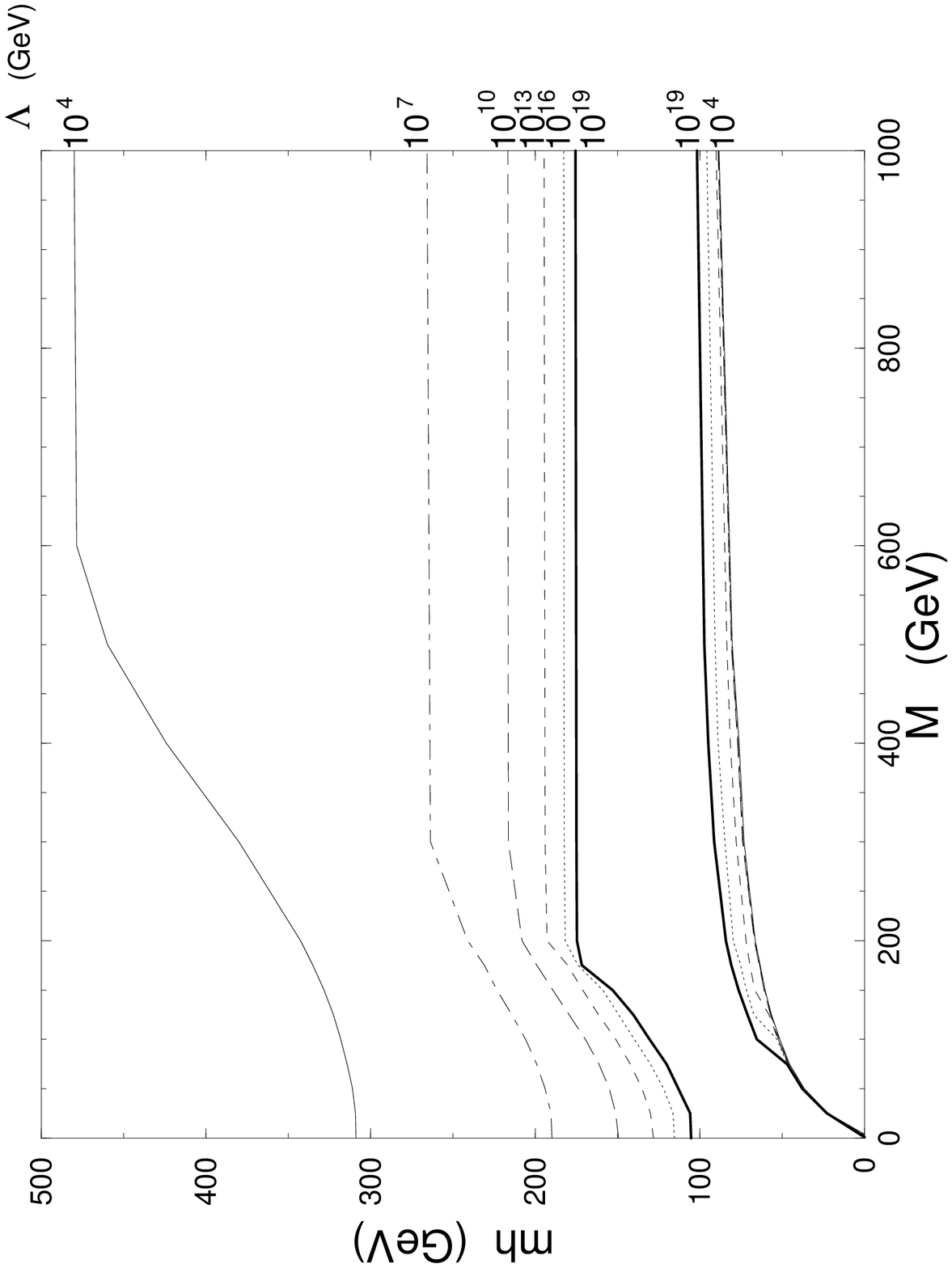}
}
\vfill
{\Large\bf Fig.~\ref{fig:t175}(a)}
\end{center}
\clearpage

\begin{center}
\makebox[0cm]{
\epsfbox{\EPSDIR 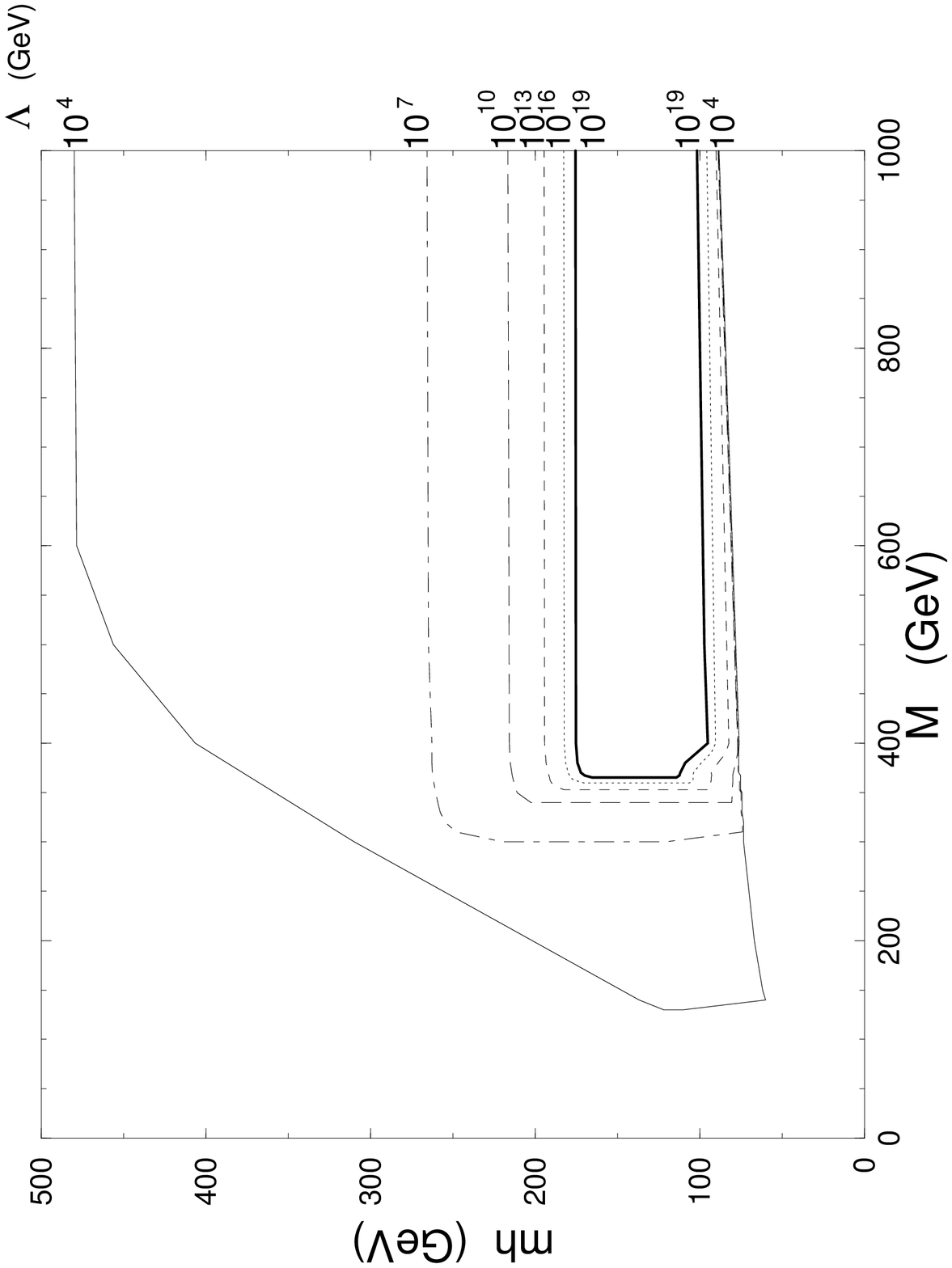}
}
\vfill
{\Large\bf Fig.~\ref{fig:t175}(b)}
\end{center}
\clearpage

\begin{center}
\makebox[0cm]{
\epsfbox{\EPSDIR 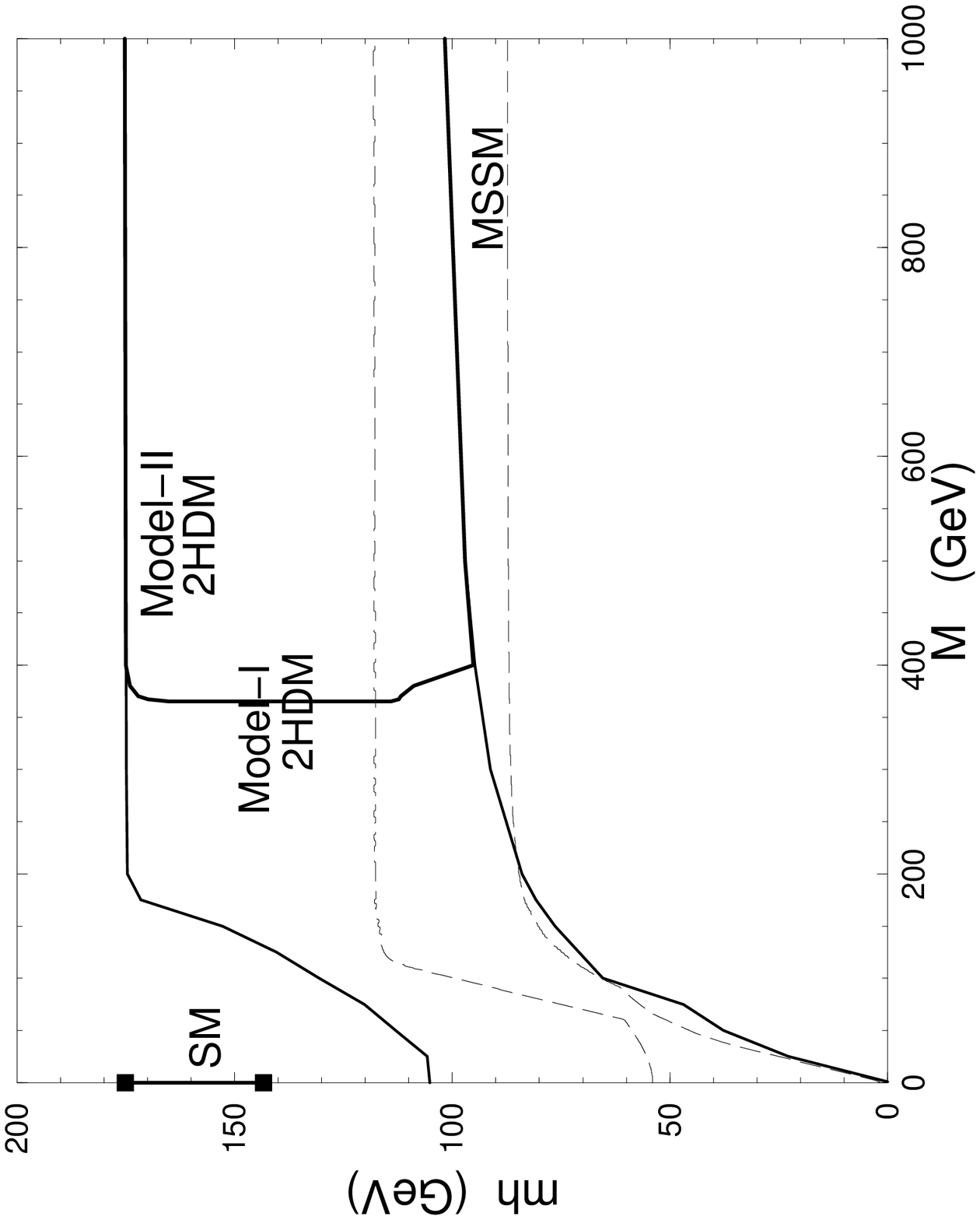}
}
\vfill
{\Large\bf Fig.~\ref{fig:combine}}
\end{center}
\clearpage

\end{document}